# Development of a cryocooler conduction-cooled 650 MHz SRF cavity operating at ~10 MV/m cw accelerating gradient

**R.C. Dhuley, S. Posen, M.I. Geelhoed, and J.C.T. Thangaraj**

Fermi National Accelerator Laboratory, PO Box 500, Batavia, Illinois, United States

Email: rdhuley@fnal.gov

**Abstract**. SRF cavities for particle acceleration are conventionally operated immersed in a bath of liquid helium at 4.2 K and below. Although this cooling configuration is practically and economically viable for large scientific accelerator installations, it may not be so for smaller accelerators intended for industrial applications such as the treatment of wastewater, sludge, flue gases, etc. In this paper, we describe a procedure to operate SRF cavities without liquid helium that can be used to construct electron-beam sources for industrial applications of electron irradiation (1-10 MeV electron energy). In this procedure, an elliptical single-cell 650 MHz niobium-tin coated niobium cavity is coupled to a closed-cycle 4 K cryocooler using high purity aluminum thermal links. The cryocooler conductively extracts heat (RF dissipation) from the cavity without requiring liquid helium around the cavity. We present construction details of this cryocooler conduction-cooling technique and systematic experiments that have demonstrated ~10 MV/m cw gradient on the cavity. By straightforward scaling up the cavity length and number of cryocoolers, the technique will provide the complete range of 1-10 MeV electron energy for industrial applications.

## 1. Introduction

Electron irradiation is a well-known method for environmental remediation processes such as the treatment of wastewater, sewage sludge, flue gases, etc. and has been demonstrated on several pilot projects [1]. To become economically viable on the large scale with existing treatment methods, electron beam (e-beam) sources capable of providing beam energy of 1–10 MeV, very high average beam power (hundreds of kW), and high wall-plug efficiency (> 50%) are needed [2]. The sources must also be robust, reliable, and have turn-key operation to be viable in harsh environments expected in industrial setting [2].

While copper based RF electrons accelerators can easily provide the required beam energy of 1-10 MeV, these must be operated in pulsed mode (RF duty cycle <<100%) to prevent overheating of the RF cavities. This is because these normal conducting copper cavities have RF surface resistance of several mΩ at room temperature and the typical GHz level RF frequencies. The pulsed mode limits the producible average e-beam power to a few tens of kW. On the other hand, superconducting radiofrequency (SRF) cavities made of niobium and $Nb_3Sn$ have surface resistance at the nΩ level at similar frequencies, which produces relatively small RF dissipation. This allows operation in continuous-wave (cw) mode, *i.e.*, with 100% RF duty cycle, which enable attaining average beam power of hundreds of kW. Superconducting RF cavities can be made with larger aperture than typical copper cavities, which is favorable for more efficient transport of high average power beams through the cavity. A meter-long SRF cavity operating at a modest 10 MV/m cw gradient can thus be a viable means to generate the required 10 MeV electron-beam energy with very high average beam power.

SRF cavities of niobium and $Nb_3Sn$, however, need ~4 K temperature to operate in the low dissipation superconducting state. In large scientific SRF accelerators the temperatures are provided by baths of liquid helium around the cavities. Although liquid helium infrastructure is economical and practical for the large, scientific accelerators (routinely several 100 m long), it can be cost-prohibitive for an industrial SRF accelerator of a meter long length. For such short-length SRF accelerators, closed-cycle 4 K cryocoolers (Pulse Tube or Gifford McMahon type) are particularly attractive because of their compact size, push-button operation, and long mean time between maintenance (>20000 hrs). Furthermore, the meter-long SRF cavity can be conductively coupled with the cryocoolers, thereby completely eliminating liquid helium from around the cavity. The resulting accelerator system is a compact, cryogenically reliable, and turn-key machine, suitable for use in industrial settings.

To demonstrate that cryocooler-cooled SRF cavities are technologically feasible, Fermilab initiated a program in 2016 to develop conduction cooling technology for SRF. In this paper, we elaborate on our developmental work and experiments that have produced an SRF cavity capable of generating the full ~10 MV/m cw accelerating gradient with conduction-cooling using a 4 K cryocooler.

## 2. Cavity preparation for cryocooler conduction-cooling

For conduction-cooled SRF cavity development we used a 650 MHz, single cell elliptical cavity made of bulk niobium that is coated with a thin $Nb_3Sn$ layer on the inner RF surface. The thermal conduction link is made using commercial sheets of high purity (5N) aluminum. A Cryomech PT420 with 2 W @ 4.2 K and 55 @ 45 K cooling capacity is selected for cooling the cavity. In this section we describe the conduction-cooling development process encompassing its conceptualization, design and simulation verification, and final fabrication.

*2.1. Conceptualization and fabrication*

Figure 1 illustrates the process of cavity preparation for conduction cooling from conceptualization to final cavity fabrication. It is well known that an elliptical cavity operating it the accelerating $TM_{010}$ mode has surface magnetic field, $H_s$ concentrated near the equator. The $H_s$ profile is shown in Figure 1a. As $H_s$ is responsible for heat dissipation, $P_{diss}$ in the cavity, the $P_{diss}$ is also expected to be concentrated near the equator. It is therefore logical to conduction-cool the cavity near the equator. Based on this rationale, a concept with flat conduction rings attached to the cavity near the equator is depicted in Figure 1a. The rings carry holes around the circumference that can be used to bolt a thermal link to the cavity.

The procedure for attaching the conduction rings to the elliptical cavity surface should (a) introduce minimal thermal resistance at the connection and (b) keep the distortion on the inner RF surface of the cavity to a minimum. To avoid use of any filler material at the connection, electron-beam welding is chosen to attach niobium conduction rings to the cavity niobium shell. Optimal weld parameters are determined using welds made on small samples. Figure 1b displays cut-section of one small sample with the curved part representing the elliptical cavity shell and the straight segment a conduction ring. The weld-parameters for this sample shows full penetration of the weld (which minimizes thermal resistance by providing maximum possible cross section) and negligible weld-bead on the opposite side of curved niobium piece (*i.e.*, the RF surface). These optimal weld-parameters are then used to weld complete conduction rings around a single cell 650 MHz niobium cavity. Figure 1c displays the fabricated cavity that has two conduction rings, welded nearly half-inch from the cavity equator. The rings have holes for bolting a thermal conduction link.

Following prerequisite RF checks on the fabricated niobium cavity, its inner (RF) surface was coated with ~2 μm layer of $Nb_3Sn$ by vapor diffusion [3]. The use of $Nb_3Sn$ on the RF surface permits cavity operation at the ~4 K temperature level in contract to pure niobium cavities that typically need ~2 K.

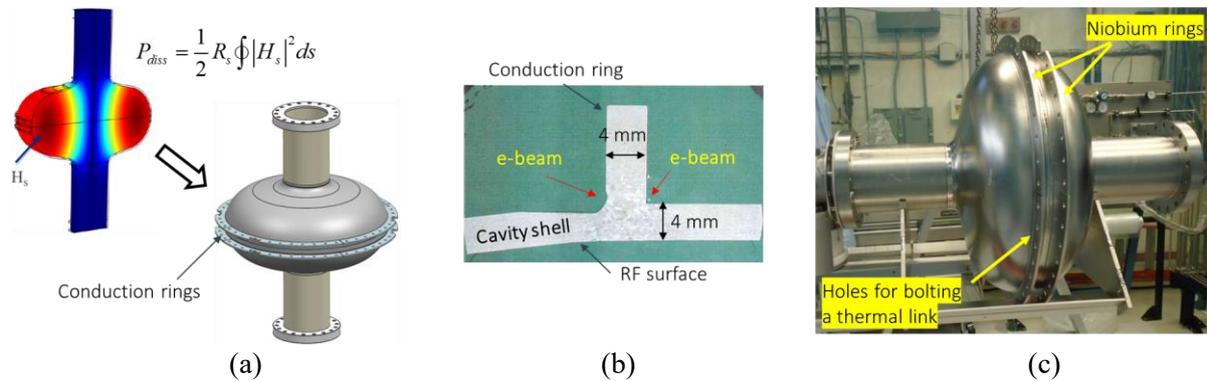

(a)                (b)                (c)

**Figure 1.** Development of the conduction-cooled SRF cavity: (left) conceptual design of conduction rings around the cavity equator, (centre) small-sample development of weld-parameters for conduction ring attachment, and (right) niobium conduction rings welded to a 650 MHz single cell cavity.

*2.2. Design and simulation verification of the thermal conduction link*

Our prior papers [4, 5] have described systemic procedures for determining thermal conductance requirements of the conduction link as well as making low thermal contact resistance bolted connections of the link with the cavity. Leveraging these, a high purity aluminum thermal link is designed. A CAD rendering of this link assembled with the cavity is shown in Figure 2a and comprises of the following: aluminum rings bolted to the niobium conduction rings on the cavity, gradual 90-degree elbows, four bus bars, a cooling distribution ring, and a flexible U-strap [6] connected to the cryocooler 4-K stage. All thermal link components are bolted together using silicon bronze screws, brass nuts, and disc springs to prevent the joints from loosening due to differential thermal contraction. All the bolted joints are interposed with 5-mil indium foil to reduce thermal contact resistance.

Following the thermal link geometric design, coupled RF and thermal simulations are performed to verify the link thermal performance. The simulations use the two boundary conditions: (a) in-house measured cryocooler capacity as a function of temperature and (b) RF heating profile of the cavity in the TM010 mode at 10 MV/m cw gradient. The material properties required for the simulations are taken from several sources: bulk thermal conductivity of niobium from [7], RF surface resistance of Nb$_3$Sn from [8]; bulk conductivity of high purity aluminum and thermal contact resistance of Al-Al and Nb-AL joints were measured in-house [5].

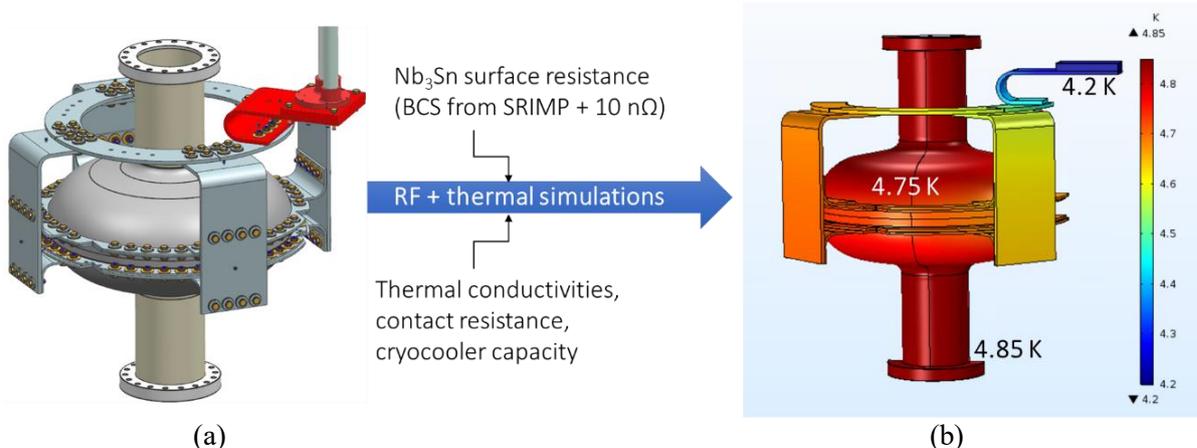

(a)                (b)

**Figure 2.** (a) CAD rendering of thermal link-cavity-cryocooler assembly and (b) simulated cavity temperature profile during 10 MV/m cw operation.

Figure 2b shows the simulated steady state surface temperature profile of the cavity. With $P_{diss}$ ≈1.9 W and the cryocooler at 4.2 K, the cavity surface remains at ≈4.8±0.05 K. The simulated thermal link conductance is ≈2.5 W/K @ ≈4.4 K.

## 3. Experimental setup

As the testing requirements and detailed description of our cavity test setup are available in [9], we only provide a brief overview of the experimental setup in this paper. Figure 3 shows a CAD rendering of our SRF cavity test setup. The $Nb_3Sn$ coated niobium cavity with the aluminum thermal links is connected to the 4-K stage of a Cryomech PT420 pulse tube cryocooler, placed on a stainless steel (SS304) plate. An MLI-wrapped Al-1100 alloy thermal shield encloses the cavity to intercept the ambient thermal radiation. The cavity as well as the thermal shield hang from the SS304 plate on 4 x 3/8" diameter low thermal conductivity titanium rods. This thermal shield is held near 50 K by conduction-cooling to the warmer stage of the cryocooler. A magnetic shield is located outside of the thermal shield. The magnetic shield is made of 2 mm thick mu-metal sheet, stays at room temperature (non-cryogenic), and provides a background total field of <15 mG at the cavity surface. A stainless steel SS304 vacuum vessel encloses all the above components and has a viton o-ring seal with the SS304 top plate.

The cavity is equipped with eight Lakeshore Cernox thermometers, four on each conduction ring, spaced at 90-degree intervals along cavity circumference. Two flux-gates are placed diametrically opposite on the conduction rings to monitor the background magnetic field during cooldown and cavity operation. A 50 W cartridge heater is provided on the thermal link immediately downstream of the cryocooler 4-K stage for controlling the cooldown rate of the cryocooler-cavity assembly. The setup uses two RF cables – an input cable for providing RF power to the cavity through one beampipe and a pick-up cable for measuring the transmitted signal from the other beampipe. All the sensor wires as well as the RF cables are thermal intercepted at the top plate of the thermal shield. The input RF cable is further intercepted at the cryocooler 4-K stage. The estimated steady heat leak to the 4-K stage is ≈0.8 W, with ≈0.5 W coming from the RF cables.

A 650 MHz, 10 W cw RF power system supplies RF power to the cavity. The computer-controlled power system measures the steady-state forward and reflected RF powers at the cavity input port as well as the transmitted signal from the cavity pick-up port. The RF system has a FPGA based closed-loop PID controller that detects any changes in the cavity frequency and quickly adjusts the source frequency to preserve resonance-locking with the cavity. The RF system also measures decay rate of the transmitted signal and derives the decay time-constant *via* an exponential curve-fit. The cavity performance parameters namely the quality factor, $Q_0$ and accelerating gradient, $E_{acc}$ can be calculated with ~10% uncertainly with the data obtained using the RF system.

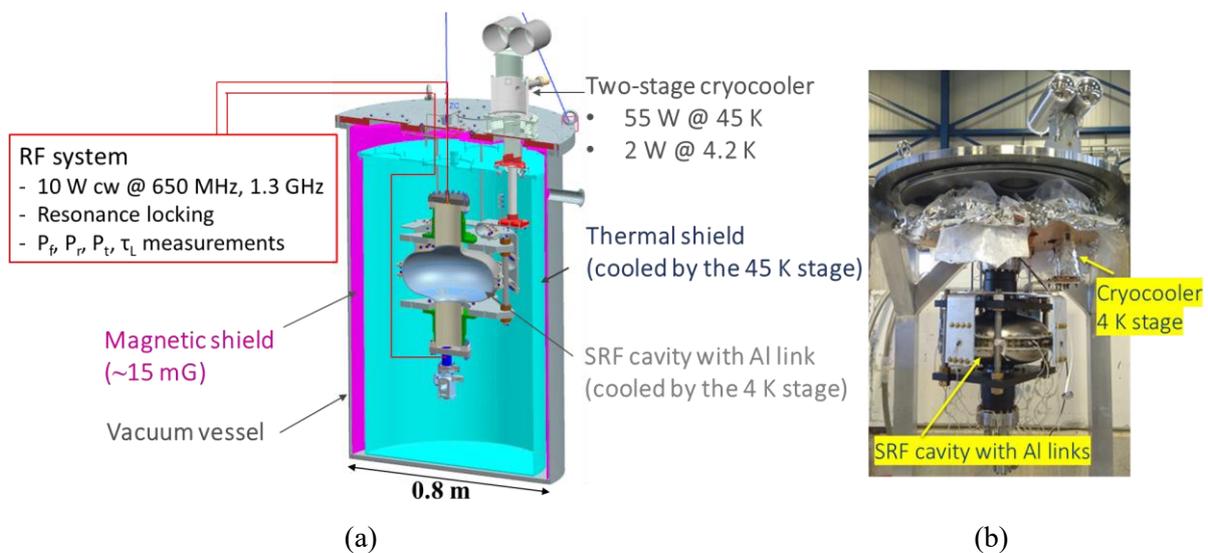

(a)          (b)

**Figure 3.** Experimental setup for conduction-cooled SRF cavity tests (a) CAD rendering (b) top assembly.

## 4. Results

*4.1. Results from initial experiments*

Baseline $Q_0$ vs. $E_{acc}$ data were first obtained on the Nb$_3$Sn coated cavity by cooling in a bath of 4.4 K liquid helium at Fermilab Vertical Test Stand (VTS). Plotted in Figure 1 as 'Coating 1, LHe', these data show that the cavity attained low field $Q_0$ = 2e10 and the target $E_{acc}$ =10 MV/m cw at $Q_0$ ≈4e9 without quench. Following this test with LHe, the cavity was warmed up and connected to the aluminum thermal links and cooled in the cryocooler test apparatus.

The first $Q_0$ vs. $E_{acc}$ data obtained using the cryocooler are plotted in Figure 5 as 'Coating 1, conduction-cooled'. The overall $Q_0$ with conduction cooling is substantially lower than the VTS test while the highest measured gradient is 6.5 MV/m cw, limited by RF power supply [10]. While investigating the cause of performance deterioration, we found that several disc springs (made of carbon steel) used on the bolted joints around the cavity had stray magnetic field >>1 G. With the cavity exposed to such high magnetic field during cooldown, the Nb$_3$Sn layer is expected to trap significant flux. It is well known that trapped field adds to the RF surface resistance and causes the $Q_0$ to drop. In contrast, as disc springs were not used during the VTS test, the cavity was exposed to no more than 10 mG field provided by the VTS magnetic shield.

*4.2. Enhancements to attain 10 MV/m cw*

Three enhancement were made to the experiment to attempt to reach the 10 MV/m cw target gradient. Firstly, all carbon steel disc springs on the bolted joints were replaced with those made of beryllium copper. This is a non-magnetic material with <0.1 mG stray field as confirmed by in-house checks. Furthermore, we ensured there are no other materials with residual/stray magnetic field inside the mu-metal magnetic shield. This ensured that the cavity surface sees <15 mG total field provided by the mu-metal shield.

Secondly, a new Nb$_3$Sn coating recipe [11] with inherently higher $Q_0$ was formulated by experimenting on another 650 MHz single cell cavity (cavity-2). After confirming that cavity-2 with new coating showed substantial improvement in the VTS, it was implemented on the cavity with conduction rings. The existing Nb$_3$Sn layer was chemically removed and then another coated using the new recipe. The new coating is referred to as 'Coating 2'.

Thirdly, we added instrumentation for controlled cooldown of the cavity-cryocooler assembly. The goal here is to do spatially uniform cavity cooldown that lower spatial temperature gradients in the cavity. By doing this, it is possible to reduce Peltier-Seebeck thermoelectric current across the Nb-Nb$_3$Sn interface. Small current implies less magnetic field generation and trapping in the Nb$_3$Sn layer, which subsequently lowers the RF surface resistance [12]. The active cooldown control is implemented by placing a 50 W cartridge heater on the thermal link immediately downstream of the cryocooled. The cavity-cryocooler assembly was first cooled down at the natural rate of the cryocooler. Spatially measured temperature at eight locations on the cavity across Nb$_3$Sn transition temperature of 18 K is plotted with time in Figure 4a showed a spatial temperature spread of >0.2 K. The cavity was then warmed to near 30 K by powering the heater on. After this, the current to the heater was lowered very slowly, which lead to more spatially uniform cooldown of the cavity. Figure 4b with temperature-time profiles with controlled cooldown shows substantially smaller spatial spread (<0.02 K) in the cavity temperature during cooldown across 18 K.

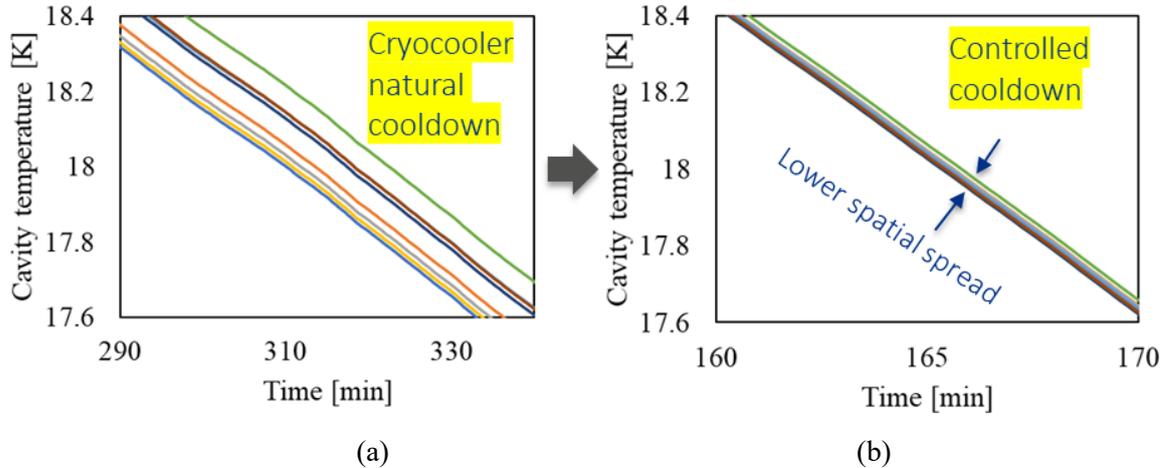

**Figure 4.** Temperature-time profiles across the Nb$_3$Sn superconducting transition temperature of 18 K measured at eight discreet locations on the cavity with (a) natural cryocooler cooldown and (b) with controlled cryocooler cooldown.

*4.3. Summary of results*

Figure 5 is a summary of $Q_0$ vs. $E_{acc}$ data of the 650 MHz Nb$_3$Sn cavity. The results for Coating 1 are described in section 4.1. The data for Coating 2 are following the enhancements presented in section 4.2. The 'Coating 2, LHe' obtained with 4.4 K liquid helium shows dramatic improvement in both $E_{acc}$ and $Q_0$ brought about by the improved Nb$_3$Sn coating recipe. The low field $Q_0$ is ~1e11, which is six times higher than the same for Coating 1. Furthermore, Coating 2 does not show a steep $Q_0$ slope like Coating 1 and crosses $E_{acc}$ of 20 MV/m cw with $Q_0$ above 1e10. This performance greatly exceeds the target set with conduction-cooling. Following this experiment with LHe, the cavity was warmed and then installed on the cryocooler test setup.

The first measurement with conduction cooling is with natural cooldown of the cryocooler, *i.e.*, starting at room temperature and allowing the cryocooler-cavity assembly to cool down to < 4 K at its natural rate. In this process, the cavity experiences substantial spatial variation in surface temperature as depicted in Figure 4a. The resulting thermocurrents are expected to trap flux in the Nb$_3$Sn layer. The data from this test, labeled as 'Coating 2, conduction-cooled, natural cooldown' in Figure 5, show some improvement over Coating 1 because of better quality of the Nb$_3$Sn layer. However, the trapped flux from thermocurrents degrades its $Q_0$ and $E_{acc}$ to values much lower than those with liquid helium. The highest cw gradient with natural cooldown of Coating 2 is measured to be 7 MV/m, limited likely by the cavity temperature of > 6 K above which we saw sudden drop in $Q_0$ (data above 7 MV/m are omitted from Figure 5).

After the test with natural cooldown, the heater on the cryocooler was power on to warm the cavity above the Nb$_3$Sn transition temperature of 18 K (normal conducting state). The current to the heater on the cryocooler was then reduced very gradually to slowly bring the Nb$_3$Sn cavity back below 18 K (superconducting state). The rate of cooldown was 5 K/hr and spatial temperature difference was <0.02 K across the 18 K transition temperature. This controlled cooldown thus kept the spatial temperature difference much smaller than during natural cooldown and is expected to substantially reduce the amount of thermocurrents induced trapped flux in the Nb$_3$Sn layer. The data labeled 'Coating 2, conduction-cooled, uniform cooldown' in Figure 5 are the $Q_0$ vs. $E_{acc}$ measured following the controlled cooldown. Low field $Q_0$ is as high as that obtained with liquid helium. This test also attained the target gradient of ~10 MV/m cw, coincidentally at a cavity temperature of ~6 K above which we again saw steep drop in $Q_0$ (data omitted).

We note that the heater ramp down rate chosen here for controlled cooldown is rather arbitrary. It will be taken as a variable parameter in future experiments to attempt to further reduce flux trapping and improve the cavity performance.

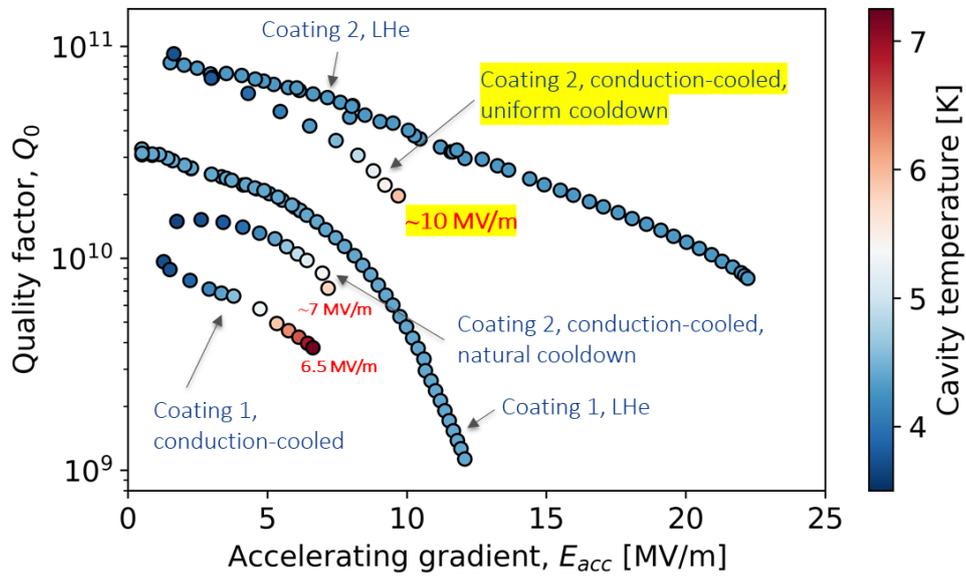

**Figure 5.** Results summary of $Q_0$ vs. $E_{acc}$ measurements on the Nb$_3$Sn coated, 650 MHz, single cell cavity.

## 5. Discussion

Having attained ~10 MV/m with conduction cooling, it is worthwhile to discuss the effectiveness of our conduction cooling technique. This is done in two ways as shown in Figure 6 and explained as follows. Figure 6a compares the experimentally determined thermal conductance of the cooling link with its simulated value. The experimental values are obtained by dividing the steady state heat flow through the link (*i.e.*, the steady state heat dissipated in the cavity) at each $E_{acc}$ with the temperature difference between the cavity conduction rings and the cryocooler. The simulated value comes from Figure 2b. The data agree reasonably to within 20%, conveying that a thermal link with predictable thermal conductance can be constructed using data obtained from our prior small sample tests [4,5,6].

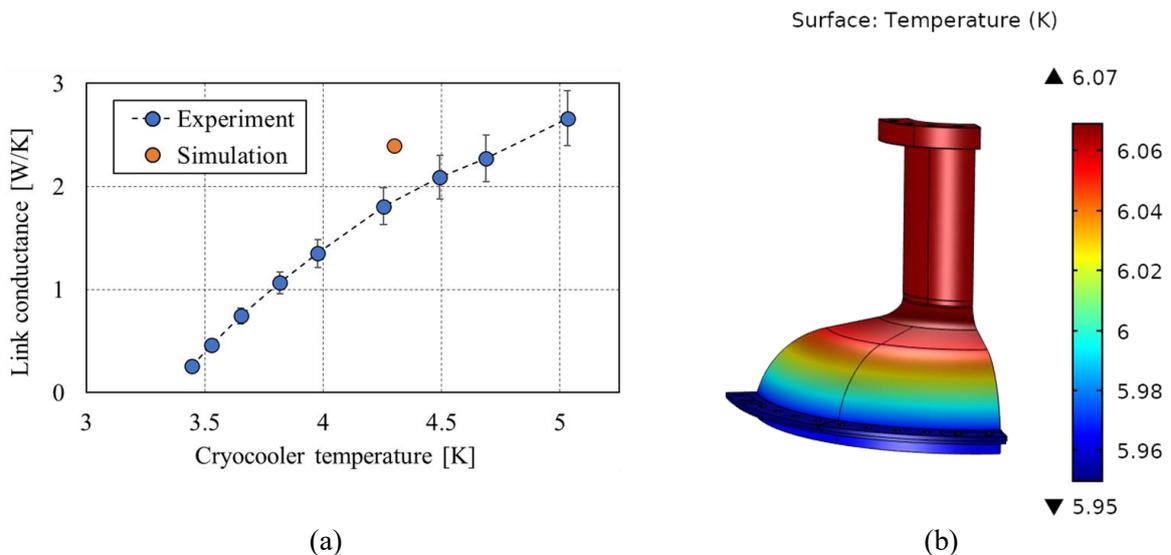

(a)  (b)

**Figure 6.** (a) Comparison of simulated and measured thermal link conductance and (b) simulated cavity surface temperature at ~10 MV/m using experimentally measured cavity temperature and RF dissipation as boundary conditions.

Figure 6b shows the surface temperature map of the cavity computed using experimentally measured temperature at the cavity rings and the RF power dissipated on the cavity inner surface during steady 10 MV/m cw operation. We highlight that the surface temperature spread from the equator (from where heat is extracted by the conduction link) to the iris <0.1 K. This implies that our technique of extracting heat from *just* near the equator during full gradient operation is effective in keeping the overall surface temperature to reasonable uniformity and so a conduction link contacting the entire surface of the cavity may not be required.

## 6. Outlook, applications, and future work

In this work, we have designed a cryocooler conduction-cooled SRF cavity and demonstrated its operation at ~10 MV/m cw gradient. We anticipate that such cryogen-free cavities will enable a new class of industrial e-beam accelerators that exploit energy-efficient SRF technology without the complexities of conventional helium cryogenic systems.

We highlight that the cryogen-free SRF cavity technique developed in the present work has already found applications in other physics research. For instance, a cryocooler-cooled conduction ring welded SRF cavity is being used for constructing a high repetition-rate field emission cathode [13] and an ultrafast electron diffraction (UED) instrument [14].

For high e-beam power accelerator applications, a few areas need further research. These include devising $Nb_3Sn$ coating procedure for multi-cell cavities (example, the standard 9-cell 1.3 GHz and 5-cell 650 MHz cavities), studying and mitigating the effects of cryocooler vibration on the cavity performance, and further optimization of the cooldown across $Nb_3Sn$ superconducting transition temperature for enhancement of cavity quality factor.

## 7. References

[1] Chmielewski A G 2011 *Reviews of Accelerator Science and Technology* **4(1)** 147-159
[2] Henderson S and Waite T 2015 Report of *Workshop on Energy and Environmental Applications of Accelerators* (Argonne, IL US Department of Energy)
[3] Posen S and Hall D L 2017 *Superconductor Science and Technology* **30** 033004
[4] Dhuley R C, Kostin R, Prokofiev O, Geelhoed M I, Nicol T H, Posen S, Thangaraj J C T, Kroc T K, Kephart R D 2019 *IEEE Transactions on Applied Superconductivity* **29(5)** 0500205
[5] Dhuley R C, Geelhoed M I, Thangaraj J C T 2018 *Cryogenics* **93** 86-93
[6] Dhuley R C, Ruschman R, Jink J T, Eyre J 2017 *Cryogenics* **86** 17-21
[7] SRIMP software available at https://www.classe.cornell.edu/~liepe/webpage/researchsrimp.html
[8] Koechlin F and Bonin B 1996 *Superconductor Science and Technology* **9** 453
[9] Dhuley R C, Geelhoed M I, Zhao Y, Terechkine I, Alvarez M, Prokofiev O, Thangaraj J C T 2020 *IOP Conf. Ser.: Mater. Sci. Eng.* **755** 012136
[10] Dhuley R C, Posen S, Geelhoed M I, Prokofiev O, Thangaraj J C T 2020 *Superconductor Science and Technology* **33** 06LT01
[11] Posen S, Lee J, Melnychuk O, Pischalnikov Y, and Seidman D 2019 Nb3Sn at Fermilab: Exploring Performance Proceedings of 19th International Conference on RF Superconductivity 818-822
[12] Stilin N, Holic A, Liepe M, Porter R, Sears J 2020 Stable CW Operation of $Nb_3Sn$ SRF Cavity at 10 MV/m using Conduction Cooling arXiv:2002.11755v1
[13] Mohsen O, Mihalcea D, Tom N, Adams N, Dhuley R C, Geelholed M I, McKeown A, Korampally V, Piot P, Salehinia I, Thangaraj J C T, Xu T 2021 *Nuclear Inst. and Methods in Physics Research A* **1005** 165414
[14] Kostin R 2020 Conduction cooled SRF photogun for UEM/UED applications *$Nb_3SnSRF'20$* https://indico.classe.cornell.edu/event/1806/contributions/1460/author/1860

FERMILAB-CONF-21-350-DI-TD

**Acknowledgments**
This manuscript has been authored by Fermi Research Alliance, LLC under Contract No. DE-AC02-07CH11359 with the U.S. Department of Energy, Office of Science, Office of High Energy Physics. Experimental work funded by U.S. Department of Energy High Energy Physics Accelerator Stewardship program. We sincerely than Dr. Roman Kostin of Euclid Techlabs, LLC for help with cavity RF and thermal simulations.